\documentclass[journal,transmag]{IEEEtran}

\usepackage{ifpdf} 

\usepackage{cite} 

\ifCLASSINFOpdf 
  \usepackage[pdftex]{graphicx} 
\else 
\fi 
\usepackage[cmex10]{amsmath} 
\usepackage{amssymb}

\usepackage{array} 
\usepackage{url} 

\usepackage{siunitx} 

\usepackage{xcolor}

\begin{document} 
%
\title{Unexpectedly allowed transition in\\two inductively coupled transmons}



\author{\IEEEauthorblockN{\'{E}tienne Dumur, 
Bruno K\"{u}ng, 
Alexey Feofanov, 
Thomas Wei{\ss}l, 
Yuriy Krupko,
Nicolas Roch, 
C\'{e}cile Naud,\\ 
Wiebke Guichard, and 
Olivier Buisson} 
\IEEEauthorblockA{Universit\'{e} Grenoble Alpes, Institut NEEL, F-38000 Grenoble, France\\ 
 CNRS, Institut NEEL, F-38000 Grenoble, France}

\thanks{Corresponding author: E. Dumur (email: etienne.dumur@neel.cnrs.fr).}}

\markboth{Journal of \LaTeX\ Class Files,~Vol.~13, No.~9, September~2014}%
{Shell \MakeLowercase{\textit{et al.}}: Bare Demo of IEEEtran.cls for Journals} 
%



\IEEEtitleabstractindextext{%
\begin{abstract} 
We present experimental results in which the unexpected zero-two transition of a circuit composed of two inductively coupled transmons is observed.
This transition shows an unusual magnetic flux dependence with a clear disappearance at zero magnetic flux.
In a transmon qubit the symmetry of the wave functions prevents this transition to occur due to selection rule.
In our circuit the Josephson effect introduces strong couplings between the two normal modes of the artificial atom.
This leads to a coherent superposition of states from the two modes enabling such transitions to occur.
\end{abstract}


\begin{IEEEkeywords} 
Josephson junctions, Cavity resonator 
\end{IEEEkeywords}}

\maketitle

\IEEEdisplaynontitleabstractindextext 

%
\IEEEpeerreviewmaketitle

\section{Introduction} 
%
%
%
%

\IEEEPARstart{C}{ircuit} Quantum ElectroDynamics (CQED) has revealed itself to be an extensive platform to address both fundamental quantum mechanics issues~\cite{654987a,sdfxcv} and applied technological interests~\cite{PhysRevX.5.021027, 987465321456a}.
In CQED, circuit elements as Josephson junctions (JJ's) and microwave resonators can be used as elementary components to engineer quantum systems such as qubits and photon cavities.
Nowadays the most commonly qubit used is the transmon qubit which consists of a small Josephson junction shunted by a capacitance~\cite{koch2007charge}.
It is described as an anharmonic oscillator with one degree of freedom.
Only the two first levels of the transmon mode have to be considered to realise a qubit.
Due to the symmetry of the wave functions, direct transition between the ground state and the second excited state of a transmon type qubit is a forbidden transition~\cite{PhysRevB.87.024510}.

In this article we present the study of a quantum system based on two inductively coupled transmons.
This circuit exhibits two degrees of freedom with a V-shape energy level diagram~\cite{PhysRevB.92.020515}.
Such an artificial atom presents a strong interest for fast qubit readout~\cite{diniz2013ultrafast}, cross-Kerr interaction~\cite{PhysRevA.84.012329,PhysRevLett.103.150503}, and single photon transistor~\cite{PhysRevLett.111.063601}.
The two degrees of freedom are given by the two normal modes of the circuit~\cite{lecocq2012coherent} which correspond to a symmetric and an antisymmetric mode.
Interestingly the symmetric mode is equivalent to the well-established transmon mode.
The Josephson nonlinearity produces an anharmonicity in the two modes of the artificial atom.
We observed by spectroscopy an unexpected transition between the ground state and the second excited state at nonzero magnetic flux in the symmetric mode.
Moreover this transition becomes forbidden close to zero magnetic flux.
These observations are discussed and explained through parity effects and the non-linear coupling of the two modes which leads to a coherent superposition of states of the symmetric and antisymmetric mode.
Selection rules and symmetry breaking has been predicted and observed in other qubits system such as coupled flux qubits~\cite{PhysRevLett.95.087001,PhysRevB.74.172505}, fluxonium~\cite{PhysRevB.87.024510,Manucharyan02102009} and flux qubit coupled to a resonator~\cite{azzaazaa}.




\section{Sample presentation}

\begin{figure}[!t] 
\centering 
\includegraphics[width=2.5in]{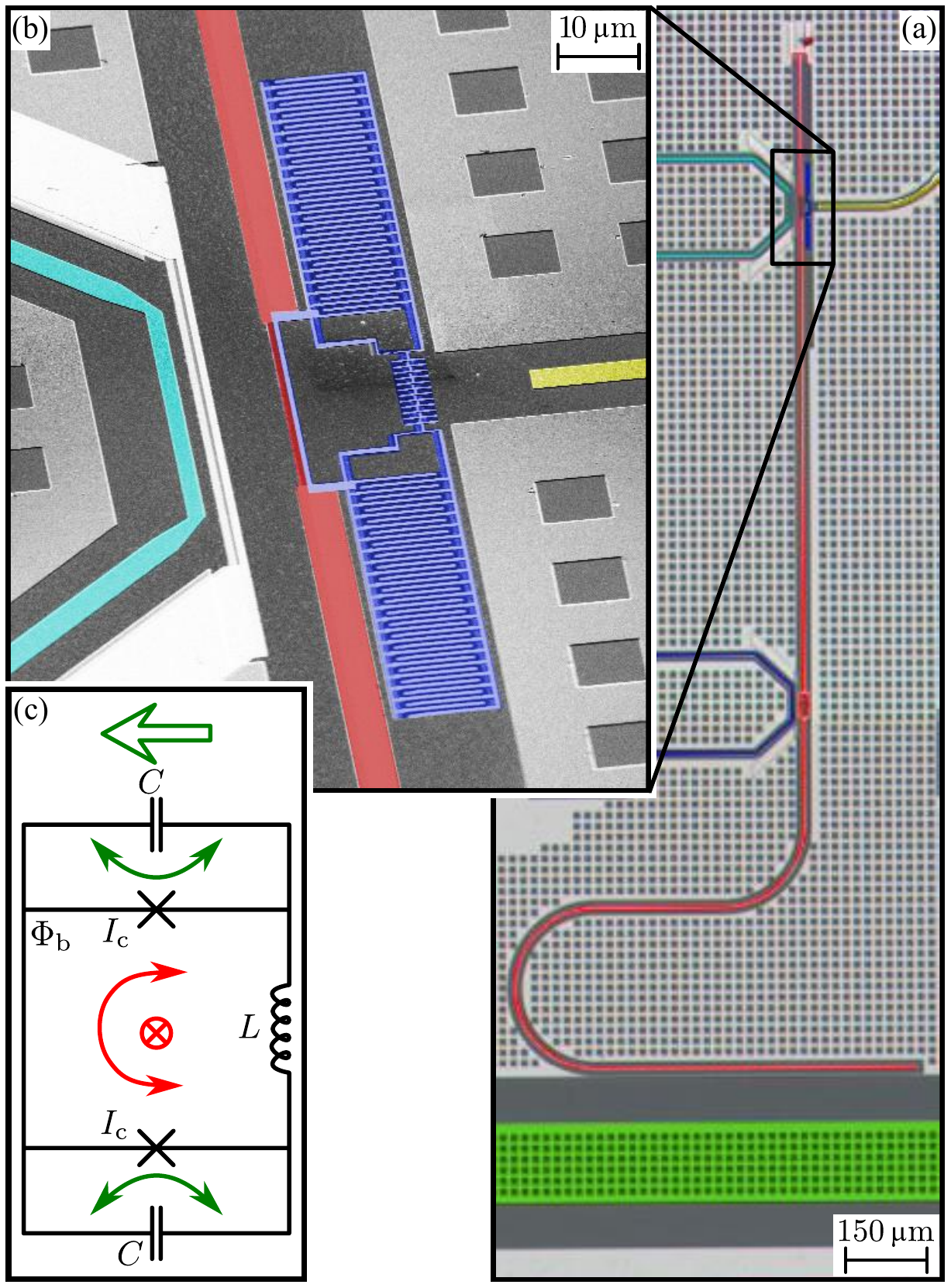} 
\caption{(a) False-colored scanning electron micrograph of the sample. The feedline (light green) through which transmission is measured is capacitively coupled to a quarter-wave resonator (red). Local DC flux bias lines (purple and cyan) allow to apply local magnetic field. The yellow line is a microwave excitation line capacitively coupled to the artificial atom. (b) Magnified view of the artificial atom. The artificial atom (blue) is capacitively and inductively coupled to microwave resonator (red). The chain of JJ is visible at the right side of the SQUID loop. (c)  Equivalent circuit diagram composed of two capacitors of capacitance $C$, two JJ's of critical current $I_\text{c}$ and an inductor of inductance $L$. The symmetric and antisymmetric modes are depicted as green and red arrows, respectively. The loop is biased with a magnetic flux $\Phi_\text{b}$.}
\label{fig:b} 
\end{figure}

Our artificial atom is composed of two identical transmons integrated into a loop of large inductance.
The transmon consists of a small JJ of critical current $I_\text{c}$ shunted by an interdigital capacitor. 
The capacitance of the JJ in parallel of the capacitor is denoted $C$. 
The dynamics of the transmon is given by the ratio of its Josephson energy $E_\text{J} = \Phi_0 I_\text{c}/(2\pi)$ on its Cooper-pair charging energy $E_\text{C} = (2e)^2/(2C)$, with $\Phi_0 = h/(2e)$ is the magnetic flux quantum. 
The linear inductance of the loop $L$ is comparable to the Josephson inductance of the transmons $L_\text{J} = \Phi_0/(2 \pi I_c)$. 
The resulting circuit is shown Fig.~\ref{fig:b}(a).

The device shown in Fig.~\ref{fig:b}(a) and Fig.~\ref{fig:b}(b) is fabricated from thin-film aluminium on a high resistive silicon substrate. 
Coarse structures are patterned by electron beam lithography and wet etched. 
Fine structure such as the artificial atom and the central line of coplanar-waveguide resonator are fabricated by lift-off using the Bridge free fabrication technique~\cite{lecocq2011junction}. 
In order to reach a linear inductance of the loop large enough to be comparable to the Josephson inductance, we incorporated in the loop a chain of 12 large JJ's \cite{Manucharyan02102009} of critical current $I_\text{c}^\prime \gg I_\text{c}$ such that $L = 12 \times \Phi_0/(2 \pi I_\text{c}^\prime)$.

A simplified diagram of our artificial atom is shown in Fig.~\ref{fig:b}(c). 
The system presents two modes of oscillations: a symmetric one corresponding to an \textit{in-phase} oscillation of the supercurrent across the two JJ and an antisymmetric one corresponding to an oscillation \textit{out-of-phase}. 
The symmetric mode can be seen as an electrical dipole pointing in line of the JJ's with an average phase $x_\text{s} = (\varphi_1 + \varphi_2)/2$ where $\varphi_{1|2}$ are the phase difference across the two JJ's.
A conjugate charge $p_\text{x} = (q_1 + q_2)/2$ is associated to this oscillating mode with $q_{1|2} = C(\Phi_0/(2\pi))\dot{\varphi}_{1|2}$ is the conjugate charge of each JJ.
The antisymmetric mode is usually not accessible in conventional SQUID due to its high frequency. 
In our artificial atom, the large inductance of the loop $L$ ensures that the frequency of the antisymmetric mode falls within our measurement bandwidth. 
That mode can be seen as a magnetic dipole pointing out of the artificial atom related to oscillations of phase difference $x_\text{a} = (\varphi_1 - \varphi_2)/2$ with a conjugate charge $p_\text{y} = (q_1 - q_2)/2$.

The artificial atom composed of two inductively coupled transmons has already been theoretically explored in greater detail in Ref. \cite{lecocq2012coherent}.
We remind here the Hamiltonian of the system written in the base of the symmetric and antisymmetric mode expanded to the fourth order by Taylor expansion
\begin{align} 
    \widehat{\mathcal{H}}   =& \;\;\;\;\; \frac{1}{2} \hbar \omega_\text{s} \left( \widehat{p}_\text{s}^2 + \widehat{x_\text{s}}^2  \right) - \hbar\omega_\text{s} \delta_\text{s} \widehat{x_\text{s}}^4  \notag \\ 
    &  +  \frac{1}{2} \hbar \omega_\text{a} \left( \widehat{p}_\text{a}^2 + \widehat{x_\text{a}}^2  \right) - \hbar\omega_\text{a} \sigma_\text{a} \widehat{x_\text{a}}^3 - \hbar\omega_\text{a} \delta_\text{a} \widehat{x_\text{a}}^4  \notag \\ 
    & + \hbar \omega_{21} \widehat{x_\text{s}}^2 \widehat{x_\text{a}} + \hbar \omega_{22} \widehat{x_\text{s}}^2 \widehat{x_\text{a}}^2, 
    \label{eq:a} 
\end{align} 
where $\widehat{x_{\text{s}|\text{a}}}$ and $\widehat{p}_{\text{s}|\text{a}}$ are the reduced position and momentum quantum operators of symmetric and antisymmetric mode such as $\widehat{x_{\text{s}|\text{a}}} = ( \widehat{a}_{\text{s}|\text{a}} + \widehat{a}_{\text{s}|\text{a}}^\dagger)/\sqrt{2}$ and $\widehat{p}_{\text{s}|\text{a}} = \mathrm{i} ( \widehat{a}_{\text{s}|\text{a}} - \widehat{a}_{\text{s}|\text{a}}^\dagger)/\sqrt{2}$. 
The first two lines of Eq.~\ref{eq:a} correspond to two independent anharmonic oscillators corresponding to the symmetric and antisymmetric mode of angular frequency $\omega_\text{s}$ and $\omega_\text{a}$ respectively.
The anharmonicity is written as dimensionless factors and is denoted $\sigma_\text{a}$ and $\delta_{\text{s}|\text{a}}$ for the corrections at the third and fourth order. 
The correction terms come from the non-linearity of the Josephson effect.
We note that the first line is equal to the Hamiltonian of the transmon qubit.
The symmetric mode is therefore similar to the usual transmon mode.
The last line shows non-linear interaction between these oscillators. 
Due to the large coupling inductance  of the loop these coupling terms have an important effect on the dynamics of the system. 
In the following, we will define $|n_\text{s}, n_\text{a} \rangle = |n_\text{s}\rangle |n_\text{a}\rangle$ as the eigenstates of the uncoupled Hamiltonian where $n_\mathrm{s|a} \in \mathbb{N}$ indexes the energy levels of each mode. 
The eigenstates of the full system will be denoted $|\psi_k\rangle$ with $k \in \mathbb{N}$ indexes the energy level

\begin{figure}[!t] 
\centering 
\includegraphics[width=\linewidth]{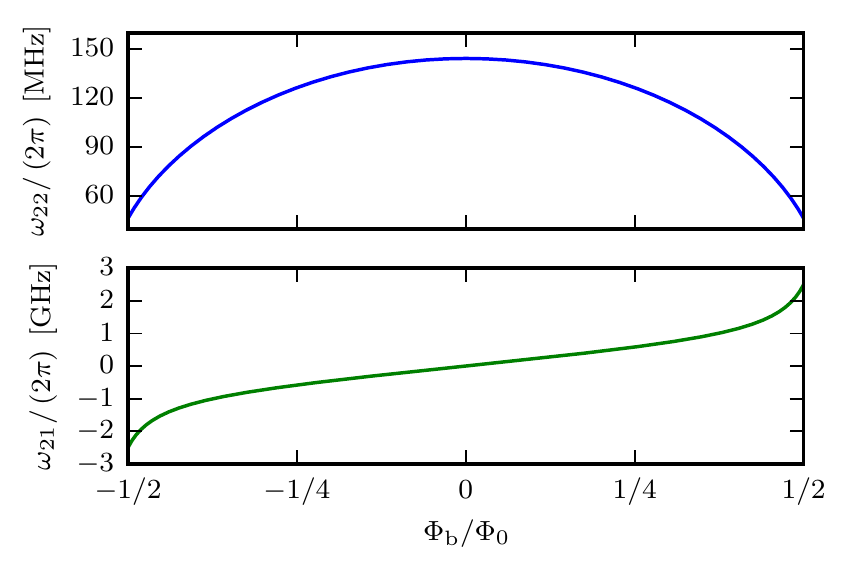} 
\caption{Dependence of the non-linear coupling terms on a magnetic flux. Typical values for $\omega_{22}/(2 \pi)$ are some hundreds of megahertz whereas $\omega_{21}/(2 \pi)$ are around few gigahertz. The theoretical predictions are realised from circuit parameters extracted from the fits performed on data presented in Fig.~\ref{fig:c}.} 
\label{fig:d} 
\end{figure} 

The non-linear coupling term $\omega_{21}$ has been studied in Ref.~\cite{lecocq2012coherent} to demonstrate at a certain flux bias a coherent frequency conversion of one excitation in the antisymmetric mode in two excitation in the symmetric one. 
In Ref.~\cite{PhysRevB.92.020515}, authors used the term $\omega_{22}$ to realise at zero flux bias a so-called V-shape energy diagram. 
This property emerges from a cross-anharmonicity between levels of the two modes. 
In Fig.~\ref{fig:d}, we show the calculated magnetic flux dependence of coupling terms. 
We see that $\omega_{21}$ exhibit a odd parity in respect to magnetic flux while $\omega_{22}$ an even one. 
This parity is related to the parity order of the Taylor expansion of the coupling term. 
We also notice the order of magnitude difference between the two couplings term.

\section{Experimental results} 

Our sample is placed in a wet dilution refrigerator with a base temperature of \SI{30}{\milli\kelvin}. 
As shown in Fig.~\ref{fig:b}(b), the artificial atom is coupled to a coplanar-waveguide resonator through a shared inductance and a stray capacitance. 
These couplings ensure that the two modes of the artificial atom will be effectively coupled to the resonator. 
By placing the circuit close to the grounded end of the resonator, we achieve a configuration in which the two coupling become comparable. 
The resonant frequency of the quarter-wave resonator is made tunable by integrating a SQUID in its central line. 
The resonant frequency can be tuned over a range of about \SI{150}{\mega\hertz} by changing the flux threading the SQUID loop. 
The readout of the artificial atom transitions is performed by standard dispersive state measurement.
The input signal is attenuated by \SI{20}{\decibel} at \SI{4.2}{\kelvin} and by \SI{40}{\decibel} at base temperature before passing through a feedline (Fig. \ref{eq:a}(a)).

Due to the capacitive coupling between the feedline and the quarter-wave resonator, the output signal carries information about the quantum state of the artificial atom.
Next, the signal is amplified by a High Electron Mobility Transistor amplifier thermalised at \SI{4.2}{\kelvin}. 
The sample is protected from noise coming from the amplifier by two isolators and a low-pass filter.

\begin{figure}[!t] 
\centering 
\includegraphics[width=\linewidth]{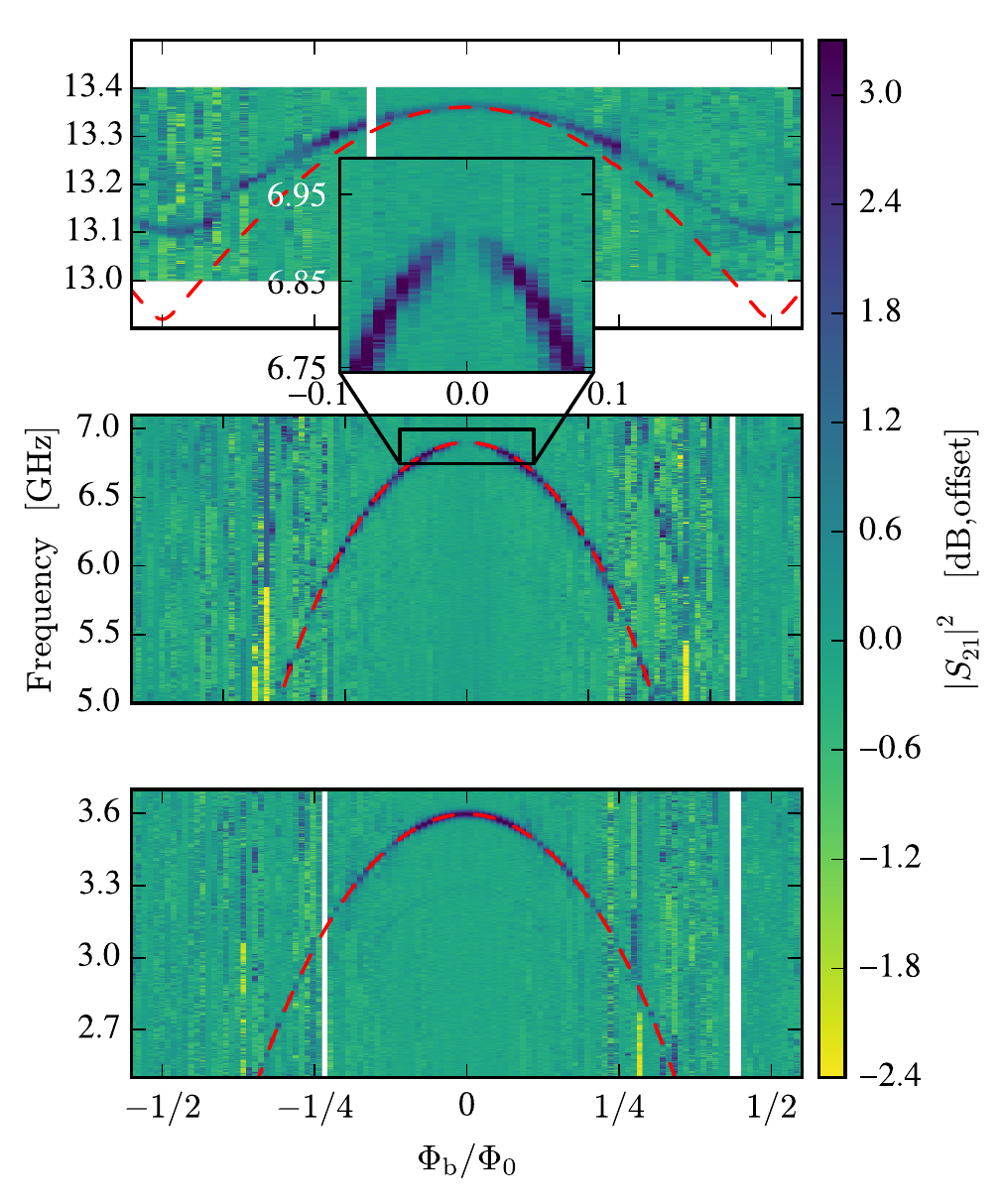} 
\caption{Spectroscopy of the artificial atom as function of frequency and magnetic field. From top to bottom, panels show the first transition of the antisymmetric mode, the second transition of the symmetric mode, and the first transition of the symmetric mode. The inset presents a zoom in the second transition of the symmetric mode for $\Phi_\text{b}/\Phi_0$ close to $0$.For each measured frequency sweep we subtracted a measurement offset. The magnetic field is converted to flux $\Phi_\text{b}$ through the SQUID loop of the artificial atom. Dashed line shows numerical model calculations of these transitions. The small discrepancy on the antisymmetric spectroscopy between experimental data and theoretical prediction close to $\Phi_\text{b}/\Phi_0 \approx \pm 1/2$ may be explained by taking into account a \SI{35}{\percent} asymmetrical critical current between the two JJ's.} 
\label{fig:c} 
\end{figure}

In the three panels presented in Fig.~\ref{fig:c}, we see three resonances of the artificial atom which depend on magnetic field. 
These curves correspond to the first and second transition of the symmetric mode and the first transition of the antisymmetric mode. 
As a function of flux, the transitions of the symmetric mode vary more strongly on a relative scale than that of the antisymmetric one. 
The antisymmetric mode involves the two JJ's as well as the linear inductance $L$ which is insensitive to flux. 
In contrast, the symmetric mode only involves the JJ's and so its transition frequencies are expected to drop to to zero as the magnetic flux tends to $\Phi_\text{b} \approx \Phi_0/2$. 
From fitting (see Ref. \cite{PhysRevB.92.020515}) we obtain the following circuit parameters $I_\text{c} = \SI{8.19}{\nano\ampere}$, $C = \SI{39.7}{\femto\farad}$, and $L = 0.192 \times L_\text{J}$.

Surprisingly in our experiments the second transition  of the symmetric mode is clearly visible.
This result was not expected since such transitions have never been observed in transmon qubits. 
Moreover, close to zero magnetic field, we observe in our experiment a disappearance of the second transition peak (see the inset in Fig.~\ref{fig:c}). 
Then we have to answer two questions: why are we able to directly measure the peak of the second level of the symmetric mode and why does the peak vanish at zero magnetic flux ?

A simple way to know whether a transition is forbidden is to look at the parity of the initial and final state as well as of the coupling operator.
In a first time we will consider the uncoupled system.
In this case, the transition probability between the ground  $|0_\text{s}\rangle |0_\text{a}\rangle$ and  the excited state $|2_\text{s}\rangle |0_\text{a}\rangle$ is given by $\mathcal{P}_{0 \rightarrow 2} \varpropto |\langle \psi_2 | \Omega_\text{s} \widehat{x}_{\text{s}} +  \Omega_\text{a} \widehat{y}_{\text{a}} | \psi_0 \rangle|^2$ where $\Omega_\text{s}$ and $\Omega_\text{a}$ are the amplitude of coupling between microwave field and the symmetric and antisymmetric mode~\cite{cohen1986mecanique}.
From previous equation we obtain $\mathcal{P}_{0 \rightarrow 2} \varpropto | \Omega_\text{s} \langle 2_\text{s} | \widehat{x}_{\text{s}} | 0_\text{s}\rangle   \langle 0_\text{a} |0_\text{a}\rangle + \Omega_\text{a} \langle 0_\text{a} | \widehat{x}_{\text{a}} | 0_\text{a}\rangle   \langle 2_\text{s} |0_\text{s}\rangle|^2 $.
With the first term we retrieve the usual results observed in transmon qubits.
Indeed the two states,  $| 0_\text{s}\rangle$ and $| 2_\text{s}\rangle$, have the same parity and the coupling term is odd. Consequently $\langle 2_\text{s}  | \widehat{x}_\text{s} | 0_\text{s} \rangle$ is zero for symmetry reason.
The second term, which does not exist in transmon type qubits, is due to the coupling between the microwave field and the antisymmetric mode.
Nevertheless due to the orthogonality of the eigenstates, this term is strictly zero.
The transition is then forbidden for the uncoupled system.


To explain why the transition is observed, we need to consider the full Hamiltonian with its non linear coupling terms.
The calculation of the eigenstates of the full Hamiltonian $| \psi_k\rangle$ is a hard problem and in our work we only consider the corrected eigenstates at the first order by quantum perturbation theory. 
The complete expression of the first order corrected eigenstates our system is given in Ref.~\cite{dumur01147222}. 
We observed that the corrected eigenstates become contaminated by states of higher and lower energy, and  more importantly, they mix states of the symmetric and antisymmetric mode due to non-linear couplings. 
By using the corrected eigenstates we derive the transition probability\footnote{The result only presents first order term, the higher order terms are neglected. 
However all terms, even those which are neglected here, exhibit the same behaviour at zero flux, they drop to zero.} $\mathcal{P}_{0 \rightarrow 2}$ which is proportional to $|(\omega_{21} \Omega_\text{a}/(8 \omega_\text{s} - 4\omega_\text{a})|^2 $.

Due to the non-linear coupling term $\omega_{21}$, the transition from the ground state to the second excited state is allowed. 
Moreover, the disappearance of the transition at zero flux can be explained by the magnetic flux dependence of $\omega_{21}$ which goes to zero at zero flux, see Fig.~\ref{fig:d}. 
We also notice that the transition is induced by the $\Omega_\text{a} \widehat{y}_{\text{a}}$ coupling operator. 
Indeed we have to keep in mind that at $\Phi_\text{b} \neq 0$, the different eigenstates of the system are a linear combination of symmetric and antisymmetric states.

\section{Conclusion} 

In this article we have presented an unexpected allowed transition between the ground and the second excited state of our artificial atom.
We also showed that this transition is allowed only at non-zero magnetic flux.
At zero magnetic flux the transition is forbidden.
The allowed/forbidden aspect of this transition as well as its magnetic flux dependence can be explained by the non-linear coupling term $\omega_{21}$ existing between the two modes of the artificial atom.
The non-linearity of this coupling is directly related to the non-linearity of the Josephson effect.
Due to this term the eigenstates of the artificial atom consist of a coherent superposition of  states from the symmetric and antisymmetric mode.
The symmetry rule, forbidding the transition in the case of the transmon qubit, no longer prevents transition in our system.
Moreover the coupling $\omega_{21}$ is magnetic flux dependent and explains the magnetic flux behaviour of the transition.
In particular at zero magnetic flux $\omega_{21}$ goes to zero, leading to a forbidden transition.

\section*{Acknowledgment} 

The authors thank  M.~Hofheinz, B.~Huard, Y.~Kubo, I.~Matei, and A.~Wallraff for fruitful discussions. 
The research has been supported by European IP SOLID and ANR-NSFC QUEXSUPERC. 
B.K. and T.W. acknowledge support from the Swiss NSF and Grenoble Nanoscience Foundation, respectively. 
W.G. is supported by Institut Universitaire de France and by the European Research Council (grant no. 306731). 
We also acknowledge the technical support of the PTA and NanoFab facilities in Grenoble.

\ifCLASSOPTIONcaptionsoff 
  \newpage 
\fi



\bibliographystyle{	} 
%

%

\begin{IEEEbiography}[{\includegraphics[width=1in,height=1.25in,clip,keepaspectratio]{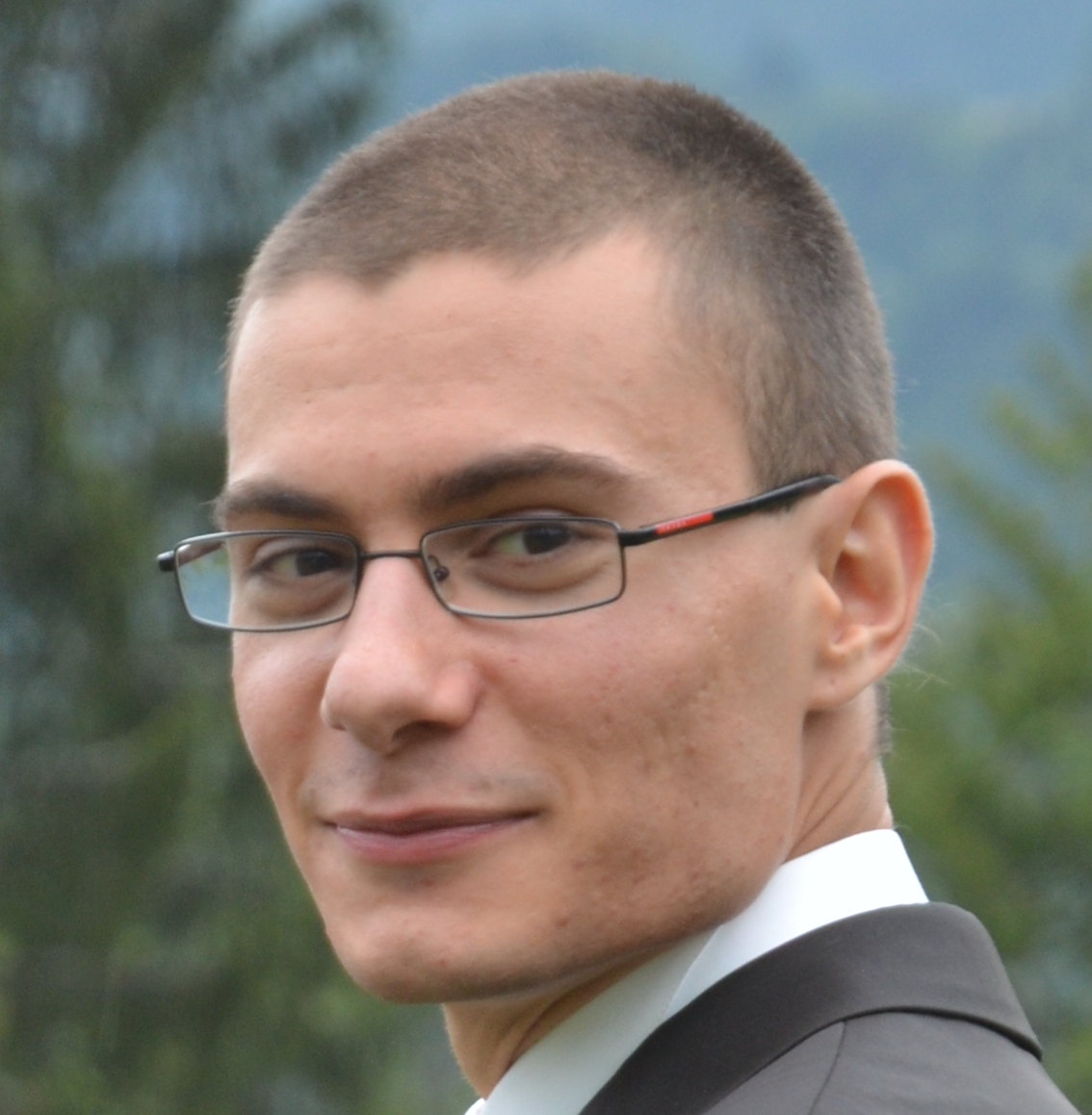}}]{\\~\\{\'E}tienne Dumur} 
was born in Champagnole, France, in 1988. He earned its master's and doctorate degrees from the University of Grenoble, France, in 2011 and 2015, respectively.
In 2011, he joined the Nano department of the N{\'e}el Institute to perform state-of-the-art experiments in Circuit Quantum ElectroDynamics.
\end{IEEEbiography} 

%
%
 



\end{document}